# Accelerating Protons to Therapeutic Energies with Ultra-Intense Ultra-Clean and Ultra-Short Laser Pulses

Stepan S. Bulanov[1], Andrei Brantov[2], Valery Yu. Bychenkov[2], Vladimir Chvykov[1], Galina Kalinchenko[1], Takeshi Matsuoka[1], Pascal Rousseau[1], Stephen Reed[1], Victor Yanovsky[1], Karl Krushelnick[1], Dale William Litzenberg[3] and Anatoly Maksimchuk[1]

[1]FOCUS Center and Center for Ultrafast Optical Science, University of Michigan, Ann Arbor, Michigan 48109

[2]P. N. Lebedev Physics Institute, Russian Academy of Sciences, Moscow 119991, Russia

[3]Department of Radiation Oncology, University of Michigan, Ann Arbor, Michigan 48109

**Abstract**

Proton acceleration by high-intensity laser pulses from ultra-thin foils for hadron therapy is discussed. With the improvement of the laser intensity contrast ratio to $10^{-11}$ achieved on Hercules laser at the University of Michigan, it became possible to attain laser-solid interactions at intensities up to $10^{22}$ W/cm$^2$ that allows an efficient regime of laser-driven ion acceleration from submicron foils. Particle-In-Cell (PIC) computer simulations of proton acceleration in the Directed Coulomb explosion regime from ultra-thin double-layer (heavy ions / light ions) foils of different thicknesses were performed under the anticipated experimental conditions for Hercules laser with pulse energies from 3 to 15 J, pulse duration of 30 fs at full width half maximum (FWHM), focused to a spot size of 0.8 microns (FWHM). In this regime heavy ions expand predominantly in the direction of laser pulse propagation enhancing the longitudinal charge separation electric field that accelerates light ions. The dependence of the maximum proton energy on the foil thickness has been found and the laser pulse characteristics have been matched with the thickness of the target to ensure the most efficient acceleration. Moreover the proton spectrum demonstrates a peaked structure at high energies, which is required for radiation therapy. 2D PIC simulations show that a 150-500 TW laser pulse is able to accelerate protons up to 100-220 MeV energies.

# I. Introduction

The nonlinear interaction of ultra-intense electromagnetic pulses, generated by compact laser systems, with plasmas has long attracted significant interest since it is accompanied by the effective conversion of laser energy into the energy of fast particles. Ion beams with a maximum energy of tens of MeV were observed in many experiments on laser pulse interaction with solid and gaseous targets [1-4]. This interaction was also thoroughly studied using 2D and 3D particle-in-cell computer simulations, which show that by optimizing the parameters of the laser pulse and the target it is possible to obtain protons with an energy of several hundreds of MeV [5-7]. The numerical and experimental studies suggest that proton therapy using compact laser systems may be practical [8-11].

Hadron therapy is a constituent part of radiation therapy, which makes use not only of high-energy ion beams but also of, electron beams, and X-rays and gamma radiation to irradiate cancer tumors (for details, see [12] and the literature cited therein). Proton therapy has a number of advantages, since one of the main challenges of radiation therapy is to deliver a desired dose to the tumor without irradiating the healthy tissues around the tumor. The proton beam is insignificantly scattered by atomic electrons and the range of protons (g/cm$^2$) with given energy is fixed, which helps to avoid undesired irradiation of healthy tissues around and behind the tumor. The presence of a sharp maximum of proton energy losses in tissues (Bragg peak) provides a substantial increase in the radiation dose in the vicinity of the beam stopping point (see [12]). Up to the present time conventional particle accelerators have been used to produce proton beams with the required parameters. The use of laser accelerators seems to be very promising because of their compactness and additional capabilities of controlling the proton beam parameters.

The acceleration of the protons to the therapeutic energy range of 200-250 MeV is the critical first step in determining whether laser acceleration of protons can be used for cancer therapy [8-10]. Proton generation in this energy range will require focusing short laser pulses to intensities of $10^{22}$ W/cm$^2$ or even higher, which is within reach of the current chirped pulse amplification (CPA) technology [13,14]. Aside from high particle energy, the therapeutic proton beam should provide a flux that is $\geq 10^{10}$ s$^{-1}$ with low energy spread of about 1%, while the typical spectrum of laser accelerated particles have Maxwellian shape with a sharp cut-off and an average energy several times less than the maximum particle energy. The selection of a narrow energy range near the end of such a spectrum would require an extremely high total number of laser triggered protons to meet radiotherapy needs. Therefore, we believe that the layered target design [15] for both increase of the number of accelerated ions and production of quasi-monoenergetic ion spectra will prevail in near future studies of high-energy proton generation for radiation therapy. Pre-plasma free interaction of the laser pulse with ultra-thin targets of solid density is necessary to achieve high-energy proton beams. However such interactions have not yet been accessible due to the low temporal intensity contrast of the existing laser systems. The temporal laser contrast, defined here as the ratio of the Amplified Spontaneous Emission (ASE) pre-pulse intensity to the peak intensity of the main pulse, must be at least 11 orders of magnitude for laser pulses with an intensity of $10^{22}$ W/cm$^2$ to avoid pre-plasma formation in the front of the target foil and preserve the physical integrity of the ultra-thin foil before the main pulse arrives.

Several "pulse cleaning" techniques have been developed for ASE suppression: saturable absorbers [16], polarization rotation [17,18], double CPA [19], cross-polarized wave generation (XPW) before pulse stretching [20], plasma mirrors [21,22], and second harmonic generation after pulse compression. One of these techniques, XPW method, was recently implemented in Hercules laser at the University of Michigan leading to a contrast ratio of $10^{-11}$ [23]. This method exploits an induced anisotropy generated by the high-intensity main peak of the linearly polarized laser pulse in cubic and tetrahedral crystals that leads to the generation of a polarized wave perpendicular to the input polarization. After passing through the crystal, the intense part of the pulse can be separated from the pre-pulse by inserting a polarizer.

In this paper we study the interaction of a super-intense, ultra-clean, ulta-short, tightly focused laser pulse with double-layer (heavy ions/ light ions) submicron foils of different thicknesses under the anticipated experimental conditions for Hercules laser in 2D PIC simulations. We show that prepulse-free 150-500 TW laser pulses focused to intensity $10^{22}$ W/cm$^2$ or even higher can produce protons with energies that are of interest for proton radiation therapy, i.e. 100-220 MeV. Simulations also indicate that the typical flux of high-energy protons is ~4 x $10^8$ per laser shot. Thus, under these conditions, it appears that a laser repetition rate of 25 Hz is required to produce flux of $10^{10}$ protons/second needed for medical applications.

The usual scenario of ion acceleration often discussed in both the experimental [1-3] and the PIC simulation [5,6,24-29] literature, is acceleration by the sheath of hot electrons generated by the high-intensity laser pulse at the front of the target. As the laser heats and accelerates these electrons they propagate through the entire target. When escaping the rear side of the foil, the electrons set up an electric field due to space-charge separation, which accelerates ions out of the target [27]. This mechanism is usually referred to as target normal sheath acceleration (TNSA).The use of ultra-thin targets in recently reported experiments [30-32] did not reveal a new mechanism of acceleration due to not sufficient laser pulse power and focused intensity. While the energy increase with the decrease of target thickness was shown, the results can be explained in the framework of

TNSA. Several other regimes of ion acceleration from thin foils were theoretically considered: Coulomb Explosion [33], the Laser Piston regime [34], enhanced TNSA [35], and Coulomb mirror [36].

In this paper, we discuss a different mechanism of ion acceleration, namely, the pre-plasma free interaction of an intense laser pulse, I~$10^{22}$ W/cm$^2$ with a solid, double-layered, sub-micron thick foil. Such a target was first proposed [8] in order to improve the quality of the accelerated proton beam. We should note that the pre-plasma free laser-solid interaction can not be achieved with standard laser intensity contrast ratios of $10^{-8}$ or worse, which do not allow direct interaction of the main laser pulse with the solid density foil at peak intensities above ~$10^{19}$ W/cm$^2$ due to the formation of a pre-plasma. Now this regime is accessible due to the contrast improvement up to $10^{-11}$ as reported in Ref. [23]. In this regime, compared to previously discussed schemes, the pulse not only expels electrons from the irradiated area but also accelerates the remaining ion core, which begins to move in the direction of pulse propagation. Then these heavy ions experience a Coulomb explosion due to the excess of positive charges, forming a strong longitudinal electric field moving with them. The light ions are accelerated in this moving electric field.

## II. Materials and Methods

In this section we describe PIC simulations for proton acceleration from thin foil targets, under the anticipated experimental conditions on the Hercules laser (3 to 15 J, a contrast ratio of $10^{-11}$, pulse duration at FWHM of 30 fs, focused to a spot size of 1 λ). The laser pulse illuminates front side of thin submicron target, which consist of electrons and heavy ions. At the rear side of target the layer of plasma with electrons and light ions – protons is attached. The principal scheme of laser interaction with double layer (heavy ions / light ions) is shown in Fig. 1.

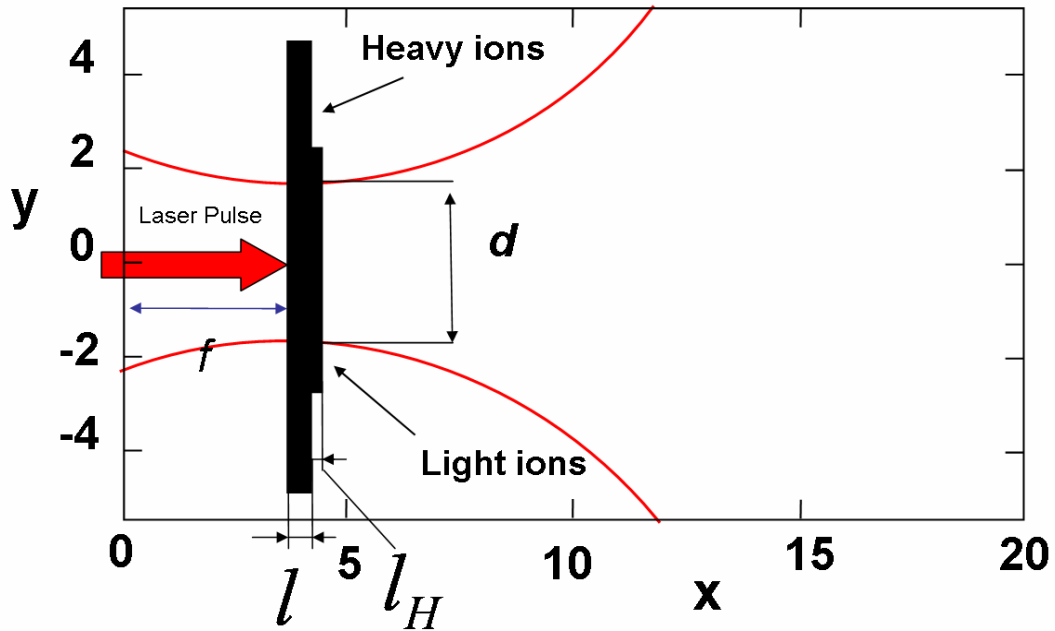

**Fig. 1 The principal scheme of laser ultra-thin double-layer (heavy ions / light ions) foil interaction. The laser pulse is focused at distance $f$ from the left border into focal spot with diameter $d$. $l$ is the thickness of heavy ion layer, $l_H$ is the thickness of light ion layer.**

## II.A. Scaling of Proton Acceleration.

When an intense laser pulse, $I \sim 10^{22}$ W/cm$^2$, directly interacts with an ultra-thin foil, it ionizes the target in a few femtoseconds maintaining the integrity of overdense plasma. Hence, the foil can be approximated as a thin layer of overdense plasma in these simulations.

It is reasonable to assume that under the action of the intense laser pulse, the electrons are evacuated from the foil region with transverse dimensions of the order of the focal spot diameter. For multi-terawatt laser pulses with femtosecond duration, the typical timescale of the hydrodynamic expansion of a submicron plasma slab is much longer than the laser pulse duration. Under these conditions heavy ions remain at rest, which results in the formation of a positively charged layer of heavy ions. However, after a time interval equal to or longer than the inverse of the heavy ion Langmuir frequency, the heavy ion layer explodes because of the repulsive Coulomb force. Moreover, if the laser field is much stronger than the Coulomb attraction field, the ions cannot retain electrons near the backside of the target, which leads to ion acceleration in the so-called Coulomb explosion regime [9,33,36].

In order to expel all the electrons and achieve Coulomb Explosion the following condition on laser electromagnetic vector-potential, $a = 0.85\ (I\ [W/cm^2]\ \lambda^2[\mu m]\ 10^{-18})^{1/2}$ and foil thickness $l$ must be satisfied

$$a > \pi\ \frac{N_e}{n_{cr}}\frac{l}{\lambda} \qquad (1)$$

Here, $N_e$ is the electron density, $n_{cr}$ is the plasma critical density, and $\lambda$ is the laser wavelength. This regime is realized for very thin foils and strong laser pulses. In order to estimate the typical energy of the accelerated ions in the Coulomb explosion regime, we assume that all the electrons produced by the ionization in the focal spot region are forced to leave the foil. In this case, the electric field near the positively charged layer is, $E_0 = 2\pi N_i Z_i e l$, where $N_i$ is the heavy ion density in the foil, and $Z_i e$ is the heavy ion electric charge. The size of the region where this estimation for $E_0$ is valid is of the order of the focal spot size, *d*, in both transverse and longitudinal directions providing one-dimensional regime of ion acceleration. While this condition persists, ion acceleration is predominantly one-dimensional. When the ions leave this region, the Coulomb explosion regime becomes three-dimensional, leading to an immediate drop in ion acceleration efficiency [37]. Thus proton layer is accelerated at the distance ~d by the electric field produced by heavy ions, $E_0 = 2\pi N_i Z_i e l$, and the maximum energy of the protons can be estimated as

$$E_{max} = \pi \ N_i Z_i e^2 l d , \qquad (2)$$

It follows from Eq. (2) that, $E_{max}$, increases linearly with foil thickness. However, the foil thickness, *l*, in equation (2) must satisfy the Coulomb explosion regime condition given by Eq. (1), $l/\lambda < a n_{cr} / \pi \ N_e$. Correspondingly, one may conclude that for given laser intensity and plasma density there is an optimum foil thickness, for which the accelerated ions reach maximum possible energy. Further increasing of foil thickness does not provide ion energy increase. In accordance with Eq. (1), for Coulomb explosion regime the optimal thickness is proportional to *a* and maximum possible proton energy scales as $E_{max} \propto a\, m\, \omega\, c\, d$.

For thicker (but still submicron) foils there is one more mechanism of ion acceleration due to light pressure. This mechanism, which works for finite reflectivity is discussed for one species ion acceleration in laser piston regime [34]. For very thin target laser light just transmits through a foil without pushing it. The transparency condition has the form [38] $l/\lambda < \pi a N_e / n_{cr}$, that coincides with Coulomb explosion condition (1). When a foil is thicker laser light reflects and accelerates foil by radiation pressure. In this case the laser field acts as a piston driving a flow of heavy ions tearing foil across. The velocity of heavy ions, $v_i$, can be estimated from momentum equation of the foil mirror

$$m_i \frac{dv_i}{dt} = \frac{1+R-T}{N_e l c} 2I, \quad (3)$$

where R and T are the target reflectivity and transmittance, and $m_i$ is the heavy ion mass. If the target thickness is the order of $l_0 = \lambda \pi a N_e / n_{cr}$ or somewhat larger the both mechanism contribute to proton acceleration. Even if electrons of the foil are not completely evacuated from the focal spot and electric field and, correspondingly, proton energy due to Coulomb explosion is reduced, the radiation pressure can compensate for this providing some proton energy increase. At the same time, for thick enough foils the radiation pressure effect will be small and one has to expect that the maximum ion energy decreases with $l$. Hence it is expected that the maximum peak proton energy as a function of foil thickness has maximum somewhere at $l > l_0$. Due to light pressure the heavy ions expand predominantly in the direction of laser pulse propagation and produce moving longitudinal charge separation electric field which accelerates the proton layer.

The final proton energy according to the acceleration mechanisms described can be estimated as

$$W = W' + v_i \sqrt{2 m_p W'}, \quad (4)$$

where $v_i$ is the heavy ions velocity defined by Eq. (3), $m_p$ is the proton mass and $W'$ is the energy that protons gain in the charge separation field, defined in the moving reference frame of the heavy ions (the maximum value of $W'$ is given by Eq. (2)).

The results of 2D PIC simulations (see section III) indicate the existence of a peak at high energies in the accelerated proton spectrum.

**II.B. 2D PIC simulations.**

In our numerical model with the 2D PIC code REMP (Relativistic ElectroMagnetic Particle, which is a mesh code based on the particle-in-cell method [39]), the acceleration of ions in high-intensity laser interactions with the thin solid-density foils is studied by using an ultra-thin two-layer aluminum-hydrogen foil. This code exploits a new scheme of current assignment that significantly reduces unphysical numerical effects of the PIC method [39]. In the simulations presented here, the uniform grid mesh size is $\lambda/200$, and the space and time scales are given in units of $\lambda$ and $2\pi/\omega$, respectively, where $\lambda$ and $\omega$, are the laser pulse wavelength and frequency respectively. The simulations are performed with 25 particles per cell.

The interaction of the laser pulse with the foil is simulated on a grid with size of $(x, y) = (20\lambda, 10\lambda)$. The laser pulse is introduced at the left boundary and propagates along the $x$-axis, from left to right. The pulse is linearly polarized along the $z$-axis. The temporal and spatial profiles of the pulse are Gaussian. The target is a double layer aluminum-hydrogen foil. The following parameters were used in simulations: laser power of 150-500 TW, pulse duration of 30 fs, and a spot size of $1.0\lambda$ (FWHM). The aluminum layer thickness was varied from $l=0.0125\lambda$ to $l=0.2\lambda$. The electron density of the foil is $400n_{cr}$, where $n_{cr}$ is the critical density of the plasma. The hydrogen layer thickness was $0.05\lambda$ with an electron density of $10n_{cr}$.

## III. Simulation Results.

In this section we present the results of the 2D PIC simulations of proton acceleration process in the pre-plasma free interaction of the laser pulse with the ultra-thin double layer foil. Since the effectiveness of proton acceleration depends not only on the laser pulse parameters but also on the foil thickness, they should be matched to each other to ensure the best acceleration regime. We first study the effect of the foil thickness on the maximum energy of protons for different laser powers (see Fig. 2).

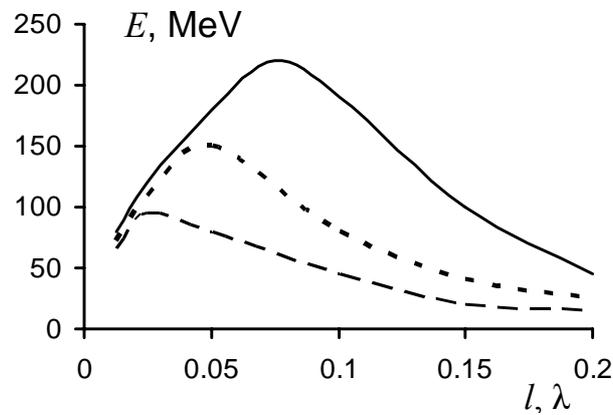

**Figure 2. The dependence of the proton maximum energy on the foil thickness for different values of the laser pulse power: 150 TW (dashed curve), 300 TW (dotted curve), and 500 TW (solid curve).**

One can clearly see that each curve has a maximum which corresponds to the optimal target thickness. The values of maximum proton energy corresponding to different laser powers almost coincide for target thickness $l < 0.05\lambda$ (see Fig. 2). The maximum energy of protons exhibits very weak dependence on the laser pulse power. It is due to the fact that for these thicknesses the condition of Coulomb Explosion is well fulfilled and the protons are accelerated by the total positive charge separation field of exploding aluminum layer that builds up after all the electrons are evacuated by the laser pulse. The strength of the Coulomb field is determined by this positive charge in the irradiated volume, which is proportional to the thickness of the foil and density of aluminum, and not by the laser pulse power.

The increase in target thickness manifests more pronounced radiation pressure effect on ion acceleration. In spite of the fact that laser pulse is no longer able to burn through the foil, it instead deposits energy into heavy ions and accelerates them by radiation pressure due to partial reflection. Then the moving heavy ion layer experiences Coulomb Explosion due to the excess of positive charge and transforms into a cloud expanding predominantly in the direction of laser pulse propagation. This expanding cloud generates a moving charge separation longitudinal electric field that accelerates the proton layer (see Fig. 3 a, b). That is why we refer to this regime of acceleration as the *Directed Coulomb Explosion* (DCE) regime. Note that radiation pressure impact on the maximum proton energy is proportional to laser intensity. So, with increase of laser intensity this effect becomes well pronounced.

In order to illustrate the process of proton acceleration in a moving longitudinal field in the DCE regime we present Fig. 3a and 3b. In Fig. 3a the distribution of electron density in the (x,y) plane at t=23 cycles is shown. In Fig. 3b the distributions of ion and proton density along with the distribution of the longitudinal component of the electric field in (x,y) plane at t=23 is shown. Here we see that the proton layer is accelerated in the moving longitudinal field, generated by the heavy ion layer expanding predominantly in the direction of laser pulse propagation.

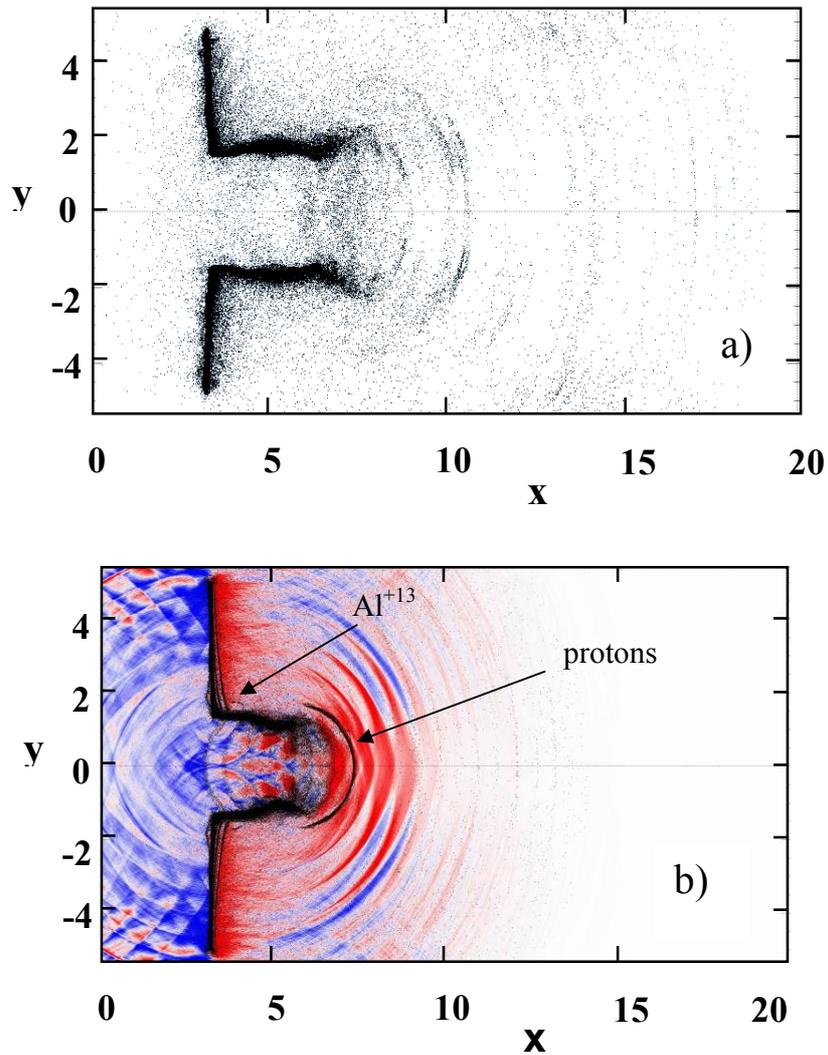

**Figure 3. (Color online) Interaction of a 500 TW laser pulse with a 100 nm thick aluminum-hydrogen foil. a) The distribution of electron density in (x,y) plane at t=23; b) The distribution of proton and $Al^{+13}$ ion density along with the distribution of longitudinal component of electric field in (x,y) plane.**

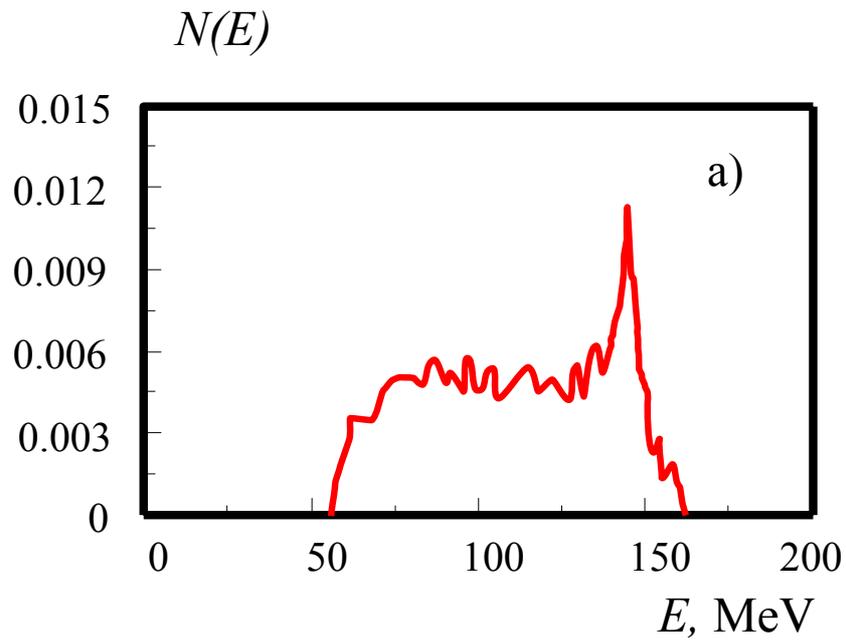

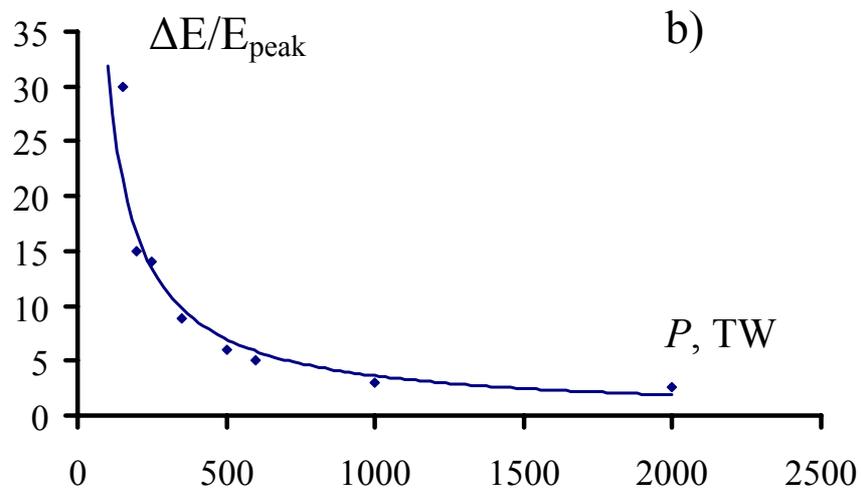

**Figure 4.** Spectrum of accelerated protons for a 500 TW laser pulse interacting with a 100 nm thick aluminum-hydrogen foil at t=36 (a). The dependence of the relative width of high energy peak on laser pulse power (b).

The spectrum of protons is presented in Fig. 4a, where one can see a peak near the energy cutoff with a width of about 10 MeV. Such a peak formation is typical for Coulomb Explosion of a target with heavy and light ions [40,41]. The width of the peak is determined by the initial thickness of the proton layer thus providing one more parameter to control the spectrum. Since the absolute width of the peak almost does not depend on laser pulse power, the relative width of the peak decreases with increase of power due to increasing of average peak energy (Fig. 4b). Moreover the advantage of using the double-layer foil is that all the protons from the focal spot are accelerated [8], so the number of accelerated particles does not depend on the laser pulse power.

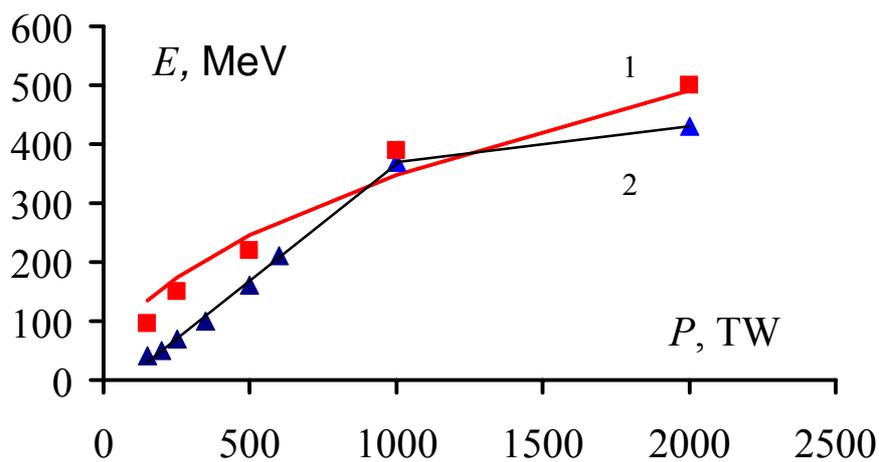

**Figure 5. (Color online)** The dependence of the maximum proton energy on the laser pulse power for optimal target thickness (squares) fitted by curve 1, and for the 0.1λ thick foil (triangles) fitted by curve2.

In Fig. 5 we present the dependence of the maximum proton energy on the laser pulse power for optimal target thickness as well as for a given foil thickness. The first curve follows a square root dependence reported before [42], while the dependence of the maximum proton energy scaling for given foil thickness is linear up to the one petawatt. Similar dependence was reported in [43], where the multiparametric PIC simulations of proton acceleration from the double-layer foil were performed. The linear dependences reflects the fact that radiation pressure prevails in proton acceleration due to laser light reflection from the foil. Increasing of laser power above one petawatt results in transparency of the foil, whose thickness is now not optimal, that changes the proton acceleration regime. The latter means that further laser power increase does not affect proton energy through Coulomb explosion since all the electrons are removed from the focal spot.

However at high values of laser power the maximum proton energy still slowly increases, indicating that another mechanism of proton acceleration comes into play. We analytically estimated the maximum energy which proton can acquire in the electrostatic field, extracted from PIC simulation. We found that proton energy should be lower by 30% than the one obtained in PIC simulations (see Fig. 6). So the electrostatic field can no longer account for proton acceleration alone. It seems that the direct acceleration of protons by the laser pulse burning through the foil can be a possible acceleration mechanism in addition to the Coulomb Explosion. That is why the proton energy is still increasing.

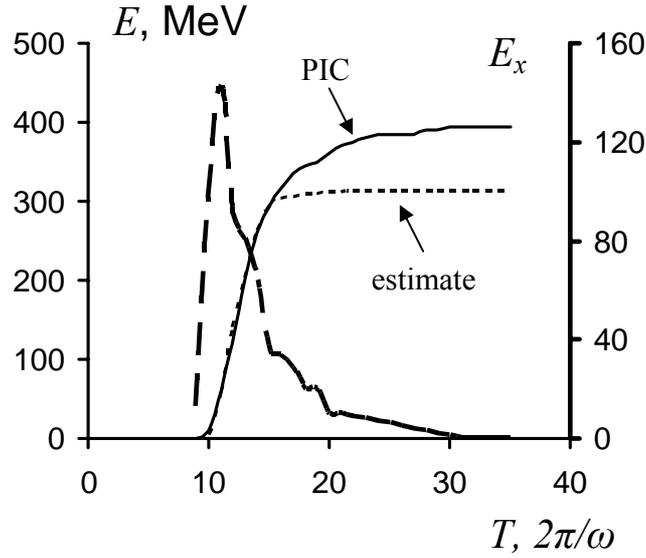

**Figure 6. The dependences of maximum proton energy (solid curve) and maximum longitudinal electric field value (dashed curve) on time in the interaction of a 1 PW laser pulse with 0.1λ double-layer foil (PIC simulation results). The estimated maximum proton energy in such longitudinal field is presented by dotted curve.**

Let's explain 30% energy gain. For this we estimate the proton energy gain which has an initial longitudinal momentum, $p_0$, in the plain electromagnetic wave with vector potential $\vec{A}$. It follows from the equations of motion that the generalized momentum is conserved, $\vec{p}_\perp - e\vec{A}/c = const$, and proton energy and longitudinal momentum are related to each other in accordance with the integral of motion : $\sqrt{m_p^2 c^4 + p_\perp^2 c^2 + p_\parallel^2 c^2} - p_\parallel c = \sqrt{m_p^2 c^4 + p_0^2 c^2} - p_0 c$, here $p_\perp$ and $p_\parallel$ are transversal and longitudinal proton momentum components in the plain wave. Taking into account the fact that $\vec{p}_\perp = e\vec{A}/c$ we get for the energy gain in the non-relativistic case

$$\frac{\Delta W}{m_p c^2} \sim \frac{a_p^2}{2}\left(1 + \frac{p_0}{m_p c}\right).$$

Here $a_p = eA/m_p c$, $m_p$ is the proton mass, $\Delta W$ is the absolute proton energy gain. We should note here that according to Lawson-Woodward theorem a free charged particle cannot gain any energy from the traveling plane electromagnetic wave over an infinite distance. However this theorem is no longer valid if the particle is injected inside the wave or interacts with a sharp (characteristic size less than a wavelength) wave front, which is the case in direct acceleration when the laser pulse burns through the target foil.

For a 1 PW laser pulse and 300 MeV protons it will give $\Delta W/W \sim 3\%$. So the acceleration cannot be described as for the plane wave and cannot account for the 30% difference between the energy gain due to longitudinal field and the total energy gain of protons. However the laser pulse burning through the foil is not a plane wave but a focused pulse, where the accelerated particle energy gain is no longer proportional to $a_p^2$, but to $a_p$ [44,45]. If we consider the energy gain of a pre-accelerated proton in a focused pulse, then

$$\frac{\Delta W}{m_p c^2} \sim a_p \left(1 + \frac{p_0}{m_p c}\right),$$

giving a 25% proton energy increase. This energy increase is in good agreement with the results of computer simulations. Thus we can conclude that for high pulse power a new mechanism of acceleration comes into play: a laser pulse that burns through the foil further accelerates the protons. However, in this case efficiency of the energy conversion to the protons is much smaller than in the case of the pure Coulomb field acceleration. So it turns out that to generate protons with higher energy it is beneficial to stay in the regime of Directed Coulomb Explosion not allowing the pulse to burn through the foil. This can be simply achieved by proper choice of foil thickness.

## IV. Discussion and Conclusions.

The investigation of multilayer targets use, as well as exploiting different regimes of laser-target interaction, is needed to the increase of the number of accelerated protons and production of quasi-monoenergetic proton spectra. The pre-plasma free interaction of a high-contrast laser pulse with ultra-thin solid targets can be a good candidate for possible laser-target design. 2D PIC simulations of such interaction were performed under the anticipated experimental conditions (3 – 15 J laser energy, 30 fs pulse duration, f/D=1.5, focusing into $10^{22}$ W/cm$^2$). The simulation demonstrated a strong dependence of the accelerated proton energy on the target thickness, indicating the existence of the optimal foil thickness which yields most energetic protons. For the optimal foil thickness, a new mechanism of acceleration comes into play, which is different from both target normal sheath and Coulomb Explosion acceleration. The laser pulse not only expels electrons from the irradiated area but also accelerates remaining heavy ions, which begin to expand due to Coulomb repulsion of excess positive charge predominantly in the direction of laser pulse propagation. The expanding heavy ion cloud generates a moving longitudinal charge separation electric field that efficiently accelerates protons from the second layer. We showed that for the anticipated experimental conditions the proton acceleration is only due to this longitudinal field. That is why we refer to this regime as the Directed Coulomb Explosion regime. We showed that it is more advantageous from the point of view of control and efficient generation of more energetic protons to employ the DCE regime, not allowing the pulse to burn through the foil. The proper matching of the target thickness to the properties of the pulse is the way to solve this problem.

The performed simulations indicate that a 500 TW laser pulse (1.0 λ FWHM) interacting with a 75 nm thick double-layered target is needed to reach a therapeutic energy of about 230 MeV (peak flux at 230 MeV of 4 x $10^8$ protons per pulse, an energy spread of 10 MeV and an emittance of $0.1\pi$ mm•mrad ). Under these conditions, 25 pulses per second would be needed to provide the necessary beam current for therapeutic applications. Moreover the advantage of using the double-layer foil is that all the protons from the focal spot are accelerated, so the number of accelerated particles does not depend on the laser pulse power, which allows the planning of the therapeutic dose. The energy spread of protons which we demonstrate is too high to meet the requirements of hadron therapy. However the use of collimators can reduce the energy spread, utilizing the fact that low energy protons have larger divergence angle. The trade off is that the number of protons will decrease, which will require higher repetition rate of the laser system.

## Acknowledgements

This study was supported by the National Science Foundation through the Frontiers in Optical and Coherent Ultrafast Science Center at the University of Michigan (PHY-0114336) , by the grant (R21 CA120262-01) from the National Institute of Health and the grant (#2289) from the International Science and Technology Center. The authors would like to thank Dr. T. Zh. Esirkepov for providing REMP code for simulations.

## References:

1. A. Maksimchuk, S. Gu, K. Flippo, D. Umstadter, and V. Y. Bychenkov, "Forward ion acceleration in thin films driven by a high-intensity laser," Physical Review Letters **84** (18), 4108-4111 (2000).

2. E. L. Clark, K. Krushelnick, J. R. Davies, M. Zepf, M. Tatarakis, F. N. Beg, A. Machacek, P. A. Norreys, M. I. K. Santala, I. Watts, and A. E. Dangor, "Measurements of Energetic Proton Transport Through Magnetized Plasma from Intense Laser Interaction with Solids," Physical Review Letters **84** (18), 670 (2000).

3. R. A. Snavely, M. H. Key, S. P. Hatchett, T. E. Cowan, M. Roth, T. W. Phillips, M. A. Stoyer, E. A. Henry, T. C. Sangster, M. S. Singh, S. C. Wilks, A. MacKinnon, A. Offenberger, D. M. Pennington, K. Yasuike, A. B. Langdon, B. F. Lasinski, J. Johnson, M. D. Perry, and E. M. Campbell, "Intense high-energy proton beams from petawatt-laser irradiation of solids," Physical Review Letters **85** (14), 2945-2948 (2000).

4. L. Willingale, S. P. D. Mangles, P. M. Nilson, R. J. Clarke, A. E. Dangor, M. C. Kaluza, S. Karsch, K. L. Lancaster, W. B. Mori, Z. Najmudin, J. Schreiber, A. G. R. Thomas, M. S. Wei, and K. Krushelnick, "Collimated Multi-MeV Ion Beams from High-Intensity Laser Interactions with Underdense Plasma," Physical Review Letters **96,** 245002 (2006).

5. T. Esirkepov, Y. Sentoku, K. Mima, K. Nishihara, F. Califano, F. Pegoraro, N. Naumova, S. Bulanov, Y. Ueshima, T. Liseikina, V. Vshivkov, and Y. Kato, "Ion acceleration by superintense laser pulses in plasmas," Journal of Experimental and Theoretical Physics Letters **70** (2), 82-89 (1999).

6. S. V. Bulanov, N. M. Naumova, T. Zh. Esirkepov, F. Califano, Y. Kato, T. V. Liseikina, K. Mima, K. Nishihara, Y. Sentoku, F. Pegoraro, H. Ruhl, and Y. Ueshima, "On the generation of collimated bunches of relativistic ions during interaction of the laser radiation with plasmas," Journal of Experimental and Theoretical Physics Letters **71** , 407 (2000).

7. Y. Sentoku, T. V. Lisseikina, T. Zh. Esirkepov, F. Califano, N. M. Naumova, Y. Ueshima, V. A. Vshivkov, Y. Kato, K. Mima, K. Nishihara, F. Pegoraro, and S. Bulanov, "High density collimated beams of relativistic ions produced by petawatt laser pulses in plasmas," Physical Review E **62**, 7271 (2000).

8. S. V. Bulanov and V. S. Khoroshkov, "Feasibility of using laser ion accelerators in proton therapy," Plasma Physics Reports **28** (5), 453-456 (2002).

9. S. V. Bulanov, T. Z. Esirkepov, V. S. Khoroshkov, A. V. Kunetsov, and F. Pegoraro, "Oncological hadrontherapy with laser ion accelerators," Physics Letters A **299** (2-3), 240-247 (2002).

10. E. Fourkal, B. Shahine, M. Ding, J. S. Li, T. Tajima, and C. M. Ma, "Particle in cell simulation of laser-accelerated proton beams for radiation therapy," Medical Physics **29** (12), 2788-2798 (2002).

11. V. Malka, S. Fritzler, E. Lefebvre, E. d'Humieres, R. Ferrand, G. Grillon, C. Albaret, S. Meyroneinc, J.-P. Chambaret, A. Antonetti, and D. Hulin, "Practicability of protontherapy using compact laser systems," Medical Physics **31** (6), 1587-1592 (2004).

12. V. S. Khoroshkov and E. I. Minakova, "Proton beams in radiotherapy" European Journal of Physics **19**, 523-536 (1998).

13. D. Strickland and G. Mourou, "Compression of amplified chirped optical pulses," Optics Communications **56** (3), 219-221 (1985)

14. S. W. Bahk, P. Rousseau, T. A. Planchon, V. Chvykov, G. Kalintchenko, A. Maksimchuk, G. A. Mourou, and V. Yanovsky, "Generation and characterization of the highest laser intensities ($10^{22}$ W/cm$^2$)," Optics Letters **29** (24), 2837 (2004).

15. H. Schwoerer, S. Pfotenhauer, O. Jäckel, K.-U. Amthor, B. Liesfeld, W. Ziegler, R. Sauerbrey, K. W. D. Ledingham, and T. Esirkepov, "Laser-plasma acceleration of quasi-monoenergetic protons from microstructured targets," Nature **439**, 445-448 (2006).

16. J. Itatani, J. Faure, M. Nantel, G. Mourou, and S. Watanabe, "Suppression of the amplified spontaneous emission in chirped-pulse-amplification lasers by clean high-energy seed-pulse injection," Optics Communications **148** (1-3), 70-74 (1998).

17. A. L. Gaeta, D. Homoelle, V. Yanovsky, and G. Mourou, "Pulse contrast enhancement of high-energy pulses by use of a gas-filled hollow waveguide," Optics Letters **27** (18), 1646 (2002).

18. A. Jullien, F. Augé-Rochereau, G. Chériaux, J.-P. Chambaret, P. d'Oliveira, T. Auguste, and F. Falcoz, "High-efficiency, simple setup for pulse cleaning at the millijoule level by nonlinear induced birefringence," Optics Letters **29** (18), 2184-2186 (2004).

19. M. P. Kalashnikov, E. Risse, H. Schonnagel, and W. Sandner, "Double chirped-pulse-amplification laser: a way to clean pulses temporally," Optics Letters **30** (8), 923-925 (2005).

20. A. Jullien, O. Albert, F. Burgy, G. Hamoniaux, J.-P. Rousseau, J.-P. Chambaret, F. Auge-Rocherau, G. Cheriaux, J. Etchepare, N. Minkovski, and S. Saltiel, "$10^{11}$ temporal contrast for femtosecond ultra-intense lasers by cross-polarizedwave generation", Optics Letters **30**, 920-922 (2005).

21. G. Doumy, F. Quere, O. Gobert, M. Perdrix, P. Martin, P. Audebert, J. C. Gauthier, J.-P. Geindre, and T. Wittmann, "Complete characterization of a plasma mirror for the production of high-contrast ultraintense laser pulses," Physical Review E, **69**(2), 26402-1-12 (2004).

22. C. Thaury, F. Quere, J.-P. Geindre, A. Levy, T. Ceccotti, P. Monot, M. Bougeard, F. Reau, P. D'Oliviera, P. Audebert, R. Marjoribanks, and Ph. Martin, "Plasma mirrors for ultrahigh-intensity optics," Nature Physics, **3**, 424-429 (2007).

23. V. Chvykov, P. Rousseau, S. Reed, G. Kalinchenko, and V. Yanovsky, "Generation of $10^{11}$ contrast 50 TW laser pulses," Optics Letters **31** (10), 1456-1458 (2006).

24. M. Allen, Y. Sentoku, P. Audebert, A. Blazevic, T. Cowan, J. Fuchs, J. C. Gauthier, M. Geissel, M. Hegelich, S. Karsch, E. Morse, P. K. Patel, and M. Roth, "Proton spectra from ultraintense laser-plasma interaction with thin foils: Experiments, theory, and simulation," Physics Of Plasmas **10** (8), 3283-3289 (2003).

25. A. Pukhov, "Three-dimensional simulations of ion acceleration from a foil irradiated by a short-pulse laser," Physical Review Letters **86** (16), 3562-3565 (2001).

26. T. Z. Esirkepov, S. V. Bulanov, K. Nishihara, T. Tajima, F. Pegoraro, V. S. Khoroshkov, K. Mima, H. Daido, Y. Kato, Y. Kitagawa, K. Nagai, and S. Sakabe, "Proposed double-layer target for the generation of high-quality laser-accelerated ion beams - art. no. 175003," Physical Review Letters **89** (17), 5003-5003 (2002).

27. S. C. Wilks, A. B. Langdon, T. E. Cowan, M. Roth, M. Singh, S. Hatchett, M. H. Key, D. Pennington, A. MacKinnon, and R. A. Snavely, "Energetic proton generationin ultra-intense laser-solid interactions," Physics of Plasmas **8** (2), 542-549 (2001).

28. M. Borghesi, J. Fuchs, S. V. Bulanov, A. J. Mackinnon, P. K. Patel, and M. Roth, "Fast ion generation by high-intensity laser irradiation of solid targets and applications," Fusion Science and Technology 49, 412-439 (2006).

29. E. d'Humieres, E. Lefebre, L. Gremillet, and V. Malka, "Proton acceleration mechanisms in high-intensity laser interaction with thin foil," Physics of Plasmas **12**, 062704 (2005).

30. D. Neely, P. Foster, and A. Robinson F. Lindau, O. Lundh, A. Persson, C.-G. Wahlström, and P. McKenna, "Enhanced proton beams from ultrathin targets driven by high contrast laser pulses," Applied Physics Letters **89**, 021502 (2006).

31. P. Antici, J. Fuchs, E. d'Humières, E. Lefebvre, M. Borghesi, E. Brambrink, C. A. Cecchetti, S. Gaillard, L. Romagnani, Y. Sentoku, T. Toncian, O. Willi, P. Audebert, and H. Pépin, "Energetic protons generated by ultrahigh contrast laser pulses interacting with ultrathin targets," Physics of Plasmas **14**, 030701 (2007).

32. T. Ceccotti, A. Levy, H. Popescu, F. Reau,1 P. D'Oliveira, P. Monot, J. P. Geindre, E. Lefebvre, and Ph. Martin, "Proton acceleration with high-intensity ultra-high contrast laser pulses," Physical Review Letters **99,** 185002 (2007).

33. E. Fourkal, I. Velchev, and C.-M. Ma, "Coulomb explosion effect and the maximum energy of protons accelerated by high power lasers," Physical Review E **71**, 036412 (2005).

34. T. Zh. Esirkepov, M. Borghesi, S. V. Bulanov, G. Mourou and T. Tajima, "Highly efficient relativistic-ion generation in the laser-piston regime," Physical Review **92**, 175003 (2004).

35. L. Yin, B. J. Albright, B. M. Hegelich, K. J. Bowers, K. A. Flippo, T. J. T. Kwan, and J. C. Fernández , "Monoenergetic and GeV ion acceleration from the laser breakout afterburner using ultrathin targets," Phys. Plasmas **14**, 056706 (2007).

36. E. Fourkal, I. Velchev, and C.-M. Ma, "Laser-induced Coulomb mirror effect: Applications for proton acceleration," Phys. Plasmas **14**, 033106 (2007).

37. V. Yu. Bychenkov and V. F. Kovalev, "On the maximum energy of ions in a disintegrating ultrathin foil irradiated by a high-power ultrashort laser pulse," Quantum Electronics **35** (12), 1143-1145 (2005).

38. G. Mourou, T. Tajima, and S. V. Bulanov, "Optics in the relativistic regime", Reviews of Modern Physics **78**, 309-371 (2006).

39. T. Zh. Esirkepov, "Exact charge conservation scheme for Particle-in-Cell simulation with arbitrary form-factor," Computer Physics Communications **135,** 144-153 (2001).

40. I. Last, I. Scheck, and J. Jortner, "Energetics and dynamics of Coulomb explosion of highly charged clusters," Journal of Chemical Physics **107**, 6685-6692 (1997).

41. I. Last and J. Jortner, "Regular multicharged transient soft matter in Coulomb explosion of heteroclusters," Proceedings of the National Academy of Sciencies of the United States of America, **102** (5), 1291-1295 (2005).

42. S. V. Bulanov, H. Daido, T. Zh. Esirkepov, V. S. Khoroshkov, J. Koga, K. Nishihara, F. Pegoraro, T. Tajima, and M. Yamagiwa, "Feasibility of Using Laser Ion Accelerators in Proton Therapy", in The Physics of Ionized Gases: 22$^{nd}$ Summer School and International Symposium on the Physics of Ionized Gases, edited by L. Hadžievski, T. Grozdanov, and N. Bibic, AIP Conference Proceedings - December 1, 2004 - Volume **740**, Issue 1, pp. 414-429.

43. T. Esirkepov, M. Yamagiwa, and T. Tajima, "Laser Ion-Acceleration Scaling Laws Seen in Multiparametric Particle-in-Cell Simulations," Physical Review Letters **96**, 105001 (2006).

44. E. Esarey, P. Sprangle, and J. Krall, "Laser acceleration of electrons in vacuum," Physical Review E **52** (5), 5443-5453 (1995).

45. Y. I. Salamin and C. H. Keitel, "Electron Acceleration by a Tightly Focused Laser Beam," Physical Review Letters **88** (9), 095005 (2002).